\documentclass[a4paper,10pt,twoside]{cpc-hepnp}
\usepackage{CJK,upgreek,fancyhdr}
\usepackage{multicol}
\usepackage{graphicx}
\usepackage{booktabs}
\usepackage{amssymb,bm,mathrsfs,bbm,amscd}
\usepackage[tbtags]{amsmath}
\usepackage{lastpage}

\begin{document}

\fancyhead[c]{\small Chinese Physics C~~~Vol. xx, No. x (201x) xxxxxx}
\fancyfoot[C]{\small 010201-\thepage}


\title{Gyroscope precession frequency analysis of a five dimensional charged rotating Kaluza-Klein black hole}

\author{%
      Mustapha Azreg-A\"{\i}nou$^{1}$
\quad Mubasher Jamil$^{2,3}$\email{mjamil@zjut.edu.cn: Corresponding Author}%
\quad Kai Lin$^{4}$
}
\maketitle

\address{%
$^1$ Engineering Faculty, Ba\c{s}kent University, Ba\u{g}l{\i}ca Campus, Ankara, Turkey\\
$^2$ Institute for Theoretical Physics and Cosmology, Zhejiang University of Technology, Hangzhou,
310023 China\\
$^3$ School of Natural Sciences, National University of Sciences and Technology,
H-12, Islamabad 44000, Pakistan\\
$^4$ Institute of Geophysics and Geomatics, China University of Geosciences,
Wuhan, Hubei 430074, China\\
}

\begin{abstract}
In this paper, we study the spin precession frequency of a test gyroscope attached to a stationary observer in the five dimensional rotating Kaluza-Klein black hole (RKKBH). We derive the conditions under which the test gyroscope moves along a timelike trajectory in this geometry and the regions where the spin precession frequency diverges. The magnitude of the gyroscope precession frequency around KK black hole diverges at two spatial locations outside the event horizon. However in the static case, the behavior of the Lense Thirring frequency of a gyroscope around KK black hole is much like an ordinary Schwarzschild black hole. Since a rotating Kaluza-Klein black hole is a generalization of Kerr-Newman black hole, we present two mass-independent schemes to distinguish these two spacetimes.
\end{abstract}

\begin{keyword}
Rotating black hole, Kaluza-Klein, spin precession, gyroscope
\end{keyword}

\begin{pacs}

\end{pacs}


\section{Introduction}
A complete gravitational collapse of a massive or a supermassive star leads to one of the following fates: neutron star, black hole or naked singularity, primarily depending on the initial mass of the star and various complicated initial conditions of physical parameters \cite{joshi}. From the mathematical and astrophysical perspectives, the task to distinguish the scenarios of the formation of black hole and naked singularities remains enigmatic. In this context we should emphasize that the recent measurement
of black hole shadow has ruled out the possibility that M87 is a naked singularity. Also all
the LIGO measurements predict a black hole-like nature of the compact objects. However from the theoretical point of view, the question remains whether a given configuration of matter collapses would lead to the formation of horizon or not. In other words, whether the horizons form prior the curvature singularity or later. Researchers pondered also on the puzzle whether a black hole can convert into a naked singularity. Numerous thought experiments involving the absorption of spinning or charged particles in an extremal black hole lead to the destruction of the horizon \cite{wald}. However the considerations of back-reaction or self-conservative force  could avoid such a conclusion \cite{cens}.

Though there are numerous  astrophysical candidates for black holes, there are none for naked singularity. It still does not discard the possibility of the existence of visible singularities since these are valid predictions of Einstein theory of General Relativity. In order to distinguish an astrophysical black hole from a naked singularity, few important schemes are proposed: the phenomenon of gravitational lensing and formation of shadow images \cite{vir}; detection of hard X and gamma rays from the inner regions of accretion disks surrounding compact objects \cite{j111}; investigation of gravitational waves emitted form compact objects \cite{kip}; and the spin precession frequency of a stationary test gyroscope being frame dragged in the ergo-region of the spinning black hole \cite{NS,j1,j2,j3,j4}. However from the theoretical perspective, only a theory of quantum gravity can ultimately explain the process of complete gravitational collapse. Observationally, the Gravity Probe B detected and measured the geodetic precession frequency of a gyroscope relative to earth \cite{B}. However, these results are not directly applicable
for black holes. Also the measurement of Gravity Probe B is a weak field result, while
we are interested in the region near the ergosphere, where strong field effects are present. The relativistic gyro-frequency diverges in the ergo-region of the black hole while for a naked singularity, it gets divergent near the singularity itself. In recent years, the gyro-frequency has been calculated for various spinning black holes and naked singularities with interesting observable consequences.

The pioneering idea of Kaluza and Klein (KK) was an attempt to unify the two fundamental forces, electromagnetism and gravity, by introducing one extra spatial dimension to an existing four dimensional spacetime structure. The hypothesis of KK asserts that the extra dimension is compactified throughout the whole spacetime so that the topology becomes that of $R^4\times S^1$. In this regard, a KK stationary solution is said to be asymptotically flat if its spacetime metric approaches a Minkowskian spacetime metric (of same dimensions) as the linear non-compactified spatial coordinates tend to infinity.

Although the approach of KK was not successful, the idea of higher dimensions has been taken over by modern string theory and M theory. In the past three decades, there have been numerous studies to derive solutions of the rotating black holes with or without electric charge in the Kaluza-Klein theory. Theoretical models of KK black holes include several fields including Maxwell field, Chern-Simons field and the dilaton field.  Besides, there are very few known black hole solutions in five dimensions, including Myers-Perry black hole \cite{mp}, Kaluza-Klein black hole with squashed horizon \cite{sq}, charged rotating black hole in minimal supergravity \cite{cho,tomi}, and in five-dimensional Einstein-Maxwell-Chern-Simons supergravity \cite{csa}. Besides there are general higher dimensional black holes in various gravitational theories such as $f(R)$ and Gauss-Bonnet with or without auxiliary fields such as electromagnetic, Yang-Mills or scalar fields \cite{mom,nie,ghoosh,ahmad,sharif,mo}. 

We take into account a stationary gyroscope moving both under the effects of relativistic frame dragging of a spinning black hole and under the effects of a constant angular speed $\Omega$ along the direction of the Killing vector $\partial_\phi$. As the spacetime is stationary, there exist another Killing vector $\partial_t$. Thus one can define a general Killing vector as $K=K^{\alpha }\partial _{\alpha }\equiv \partial _t+\Omega\partial_{\phi }$, which is a linear combination of the two Killing vectors $\partial_t$ and $\partial_\phi$, \emph{provided $\Omega$ is a constant}, that is, for $\partial _t+\Omega\partial_{\phi }$ to be a Killing vector, $\Omega$ has to be a constant. We consider the case where the actual velocity $u$ of the gyroscope is proportional to $K$: $u=|K|/\sqrt{|K^2|}$. Note that $g_{tt}=0$ gives information about the stationary limit surfaces or the ergo-regions in the spacetime. In the limit $\Omega\rightarrow 0$, one recovers the expression of Lense-Thirring precession frequency. The geodetic precession effect of a parallel- transported spin vector along a circular geodesic in five-dimensional squashed Kaluza-Klein black hole spacetime has already been investigated \cite{already}. For numerous advances in the studies of gyroscopic precession frequency in various gravitational theories and different geometric spacetimes, the interested reader might refer to \cite{a}.

The plan of the paper is as follows: In Sec. II, we briefly provide a review of Kaluza-Klein theory. In Sec. III, we present a brief review with new physical insights of the rotating KKBHs. Next we calculate the general spin precession frequency vector of the gyroscope around RKKBH and discuss some physical consequences in Sec. IV. In Sec. V,  we develop the general formalism of gyroscope spin precession frequency for a rotating black hole in five dimensional Kaluza-Klein theory. In Sec. VI, we discuss how to distinguish RKKBH from the ordinary Kerr-Newman black hole (KNBH) using spin precession analysis. Finally, we conclude in Sec VII. We have added an appendix section to discuss technical matters pertinent to Sec. IV.\\

\section{Brief Review of Kaluza-Klein Theory}

In this work we adopt the following index conventions, most for KK theories. ($\alpha,\beta,\gamma,\delta$): $1\to 5$, ($\mu,\nu,\rho,\sigma$): $1\to 4$, ($i,j,k,l,m,n$): $1\to 3$, and ($a,b,c,d$): $1,\; 2$. We work with the general metric ansatz for a five-dimensional spacetime: $x^i \text{ (spatial dimensions)},\,x^4=t,\,x^5=\psi$ (fifth or extra dimension).

Kaluza and Klein studied the Einstein's theory of general relativity in five dimensions in order to unify the gravitation with electromagnetism (see~\cite{kkk} for a review). They assumed that the five dimensional universe is empty and satisfies the field equations:
\begin{equation} \label{G1}
\hat G_{\alpha\beta}=0,~~ \hat R_{\alpha\beta}=0,
\end{equation}
which could be derived from the corresponding five dimensional action
\begin{equation}\label{ss1}
\hat S=-\frac{1}{16\pi \hat G}\int \hat R \sqrt{-\hat g}\,{\rm d}^4x {\rm d}\psi,
\end{equation}
Notice that the definitions of Christoffel symbol, Ricci scalar and Einstein tensor are identical to those of four dimensions. In order to incorporate the electromagnetism $A_\alpha$ along with gravity $g_{\mu\nu}$, Kaluza introduced one more scalar $\varphi$ and consequently proposed to decompose the five-dimensional metric in the form
\begin{equation}\label{g1}
\hat g_{\alpha\beta}=\begin{pmatrix}
g_{\mu\nu}+\kappa^2 \varphi^2 A_\mu A_\nu & \kappa \varphi^2 A_\mu \\
\kappa \varphi^2 A_\nu & \varphi^2
\end{pmatrix}
\end{equation}
where $\kappa^2=16\pi G$. By substitution of the metric (\ref{g1}) in (\ref{G1}) yields the following field equations in four dimensions:
\begin{eqnarray}\label{eqns}
G_{\mu\nu}=\frac{\kappa^2\varphi^2}{2}T^{EM}_{\mu\nu}-\frac{1}{\varphi}[\nabla_\mu\nabla_\nu\varphi-g_{\mu\nu}\Box\varphi],~~ \nabla^\mu F_{\mu\nu}=-3\frac{\nabla^\mu\varphi}{\varphi}F_{\mu\nu}, ~~\Box\varphi=\frac{\kappa^2\varphi^3}{4}F_{\mu\nu}F^{\mu\nu},
\end{eqnarray}
which is a set of fifteen equations with fifteen unknowns (i.e. $10,\,4,\,1$ components of $g_{\mu\nu}$, $A_\mu$ and $\varphi$ respectively). It is interesting to note that the above set of field equations can also be derived by the variation of the following four dimensional action
\begin{equation}\label{oct}
S=\int {\rm d}^4x \sqrt{-g}\varphi\Big( \frac{R}{\kappa^2}+\frac{1}{4}\varphi^2F_{\mu\nu}F^{\mu\nu}+\frac{2}{3\kappa^2}\frac{\nabla^\mu\varphi\nabla_\mu\varphi}{\varphi^2} \Big),
\end{equation}
which is presently an action in the Jordan frame. Note that the field equations Eqs. (\ref{eqns}) have been derived from the action Eq. (\ref{oct}). In order to recast the last action in the Einstein frame, we employ a conformal transformation $g_{\alpha\beta}\rightarrow g'_{\alpha\beta}=\Omega^2 g_{\alpha\beta}$, and further replacing $\varphi^2\rightarrow\varphi$ and $\Omega^2\rightarrow \varphi^{-1/3}$ in (\ref{oct}), we obtain
\begin{equation}
S'=\int {\rm d}^4x \sqrt{-g'}\Big( \frac{R'}{\kappa^2}+\frac{1}{4}\varphi F_{\mu\nu}F^{\mu\beta}+\frac{1}{6\kappa^2}\frac{\nabla^\mu\varphi\nabla_\mu\varphi}{\varphi^2} \Big).
\end{equation}
Moreover, if we substitute a dilaton field $\sigma\equiv \frac{1}{\sqrt{3}\kappa}\ln\varphi$, we obtain the canonical form of the four dimensional action in the Einstein frame as follows:
\begin{equation}
S'=\int {\rm d}^4x \sqrt{-g'}\Big( \frac{R'}{\kappa^2}+\frac{1}{4}{\rm e}^{\sqrt{3}\kappa \sigma}F_{\mu\nu}F^{\mu\nu}+\frac{1}{2}\nabla^\mu\sigma\nabla_\mu\sigma \Big).
\end{equation}
The last action describes a dilaton scalar field coupled to both gravity and electromagnetism. If there is no electromagnetism involved, then the last action describes a scalar field minimally coupled to gravity with no potential.

The Gibbons-Hawking-York (GHY) boundary term is added only in case the manifold $M$ has a boundary, which is a 3-dimensional hypersurface denoted usually by $\partial M$. The field equations are the same whether the manifold has a boundary or not. If $\partial M$ exists one sets $\delta g_{\alpha\beta}=0$ on the boundary $\partial M$ as a further constraint in order to obtain the same field equations one should obtain if no boundary at all. In the present case, there is no boundary involved and consequently there is no need to account for the GHY boundary term as was done in~\cite{nuc}.

\section{Rotating Kaluza-Klein Black Hole}

Static Kaluza-Klein black holes are derived by standard methods of solving the Einstein field equations or Einstein-Yang-Mill equations with matter fields \cite{radu}. However, the rotating Kaluza-Klein black holes are not, in general, derived by solving the field equations. Instead one employs the product of Kerr metric with a line, boosts along the line and then compactifies the extra dimension~\cite{horo,horo2}, see also \cite{sheykhi,nuc} where the solution is derived by solving the Einstein-Maxwell and scalar field equations. The resulting solution is stationary, axis-symmetric and invariant under translation along the fifth dimension. Motivated by higher dimensional string and supergravity theories, researchers have derived six and multi-dimensional rotating Kaluza-Klein black holes as well \cite{high}. 

The rotating black hole in the Kaluza-Klein theory (RKKBH) is given in the form~\cite{npb}:
\begin{equation}\label{solution}
	{\rm d}s^2=\frac{H_2}{H_1}({\rm d}\psi +A)^2-\frac{H_3}{H_2} ({\rm d}t+B)^2+H_1 \Big(  \frac{{\rm d}r^2}{\Delta}+{\rm d}\theta^2+\frac{\Delta}{H_3}\sin^2\theta {\rm d}\phi^2 \Big),
\end{equation}
where
\begin{eqnarray}
	H_1 &=& r^2 +a^2\cos^2\theta +r(p-2m)+\frac{p(p-2m)(q-2m)}{2(p+q)}-\frac{p}{2m(p+q)}\sqrt{(q^2-4m^2)(p^2-4m^2)}~a\cos\theta,\nonumber\\
	H_2 &=& r^2 +a^2\cos^2\theta +r(q-2m)+\frac{q(p-2m)(q-2m)}{2(p+q)}+\frac{q}{2m(p+q)}\sqrt{(q^2-4m^2)(p^2-4m^2)}~a\cos\theta,\nonumber\\
	H_3 &=& r^2+a^2\cos^2\theta-2mr,\qquad \Delta = r^2+a^2-2mr,\nonumber\\
	\nonumber
\end{eqnarray}
including the one-forms
\begin{eqnarray}
	A&=& - \frac{1}{H_2}\Big[  2Q(r+\frac{p-2m}{2})+\sqrt{\frac{q^3(p^2-4m^2)}{4m^2 (p+q)}}a\cos\theta  \Big]{\rm d}t
	-\frac{1}{H_2}\Big[  2P(H_2+a^2\sin^2\theta)\cos\theta+\sqrt{\frac{p(q^2-4m^2)}{4m^2 (p+q)^3}}\nonumber\\
	&&\times[(p+q)(pr-m(p-2m))+q(p^2-4m^2)]a\sin^2\theta  \Big]{\rm d}\phi,\\
	&\equiv &A_4{\rm d}t+A_3{\rm d}\phi,\nonumber\\
	\label{B3}B&=&\frac{(pq+4m^2)r-m(p-2m)(q-2m)}{2m(p+q)H_3}\sqrt{pq}a\sin^2\theta {\rm d}\phi,\\
	&\equiv &B_3{\rm d}\phi .\nonumber
\end{eqnarray}
The black hole metric given by Eq. (\ref{solution}) is a solution of the field equations Eqs. (\ref{eqns}) which are derived from the action Eq. (\ref{oct}). The four parameters $m,\,a,\,p,\,q$ appearing in the solution are related to the physical mass $M$, angular momentum $J$, electric charge $Q$ and magnetic charge $P$ as follows:
\begin{equation}\label{para}
M=\frac{p+q}{4},~~ J=\frac{\sqrt{pq}(pq+4m^2)}{4m(p+q)}~a,~~ Q^2=\frac{q(q^2-4m^2)}{4(p+q)},~~ P^2=\frac{p(p^2-4m^2)}{4(p+q)}.
\end{equation}
One may reverse these formulas to express $m,\,a,\,p,\,q$ in terms of $M,\,J,\,P,\,Q$, however, the obtained expressions are sizable and we will not derive them. The detailed procedure for deriving them is described in the paragraph including Eqs.~\eqref{eta2} through~\eqref{bc}. 

The corresponding four dimensional metric in the coordinates $(t,r,\theta,\phi)$ in the Einstein frame is
\begin{equation}\label{met}
	{\rm d}\bar{s}^2=-\frac{H_3}{\rho^2}{\rm d}t^2-2\frac{H_4}{\rho^2}{\rm d}t {\rm d}\phi+\frac{\rho^2}{\Delta} {\rm d}r^2+\rho^2 {\rm d}\theta^2+
	\Big(  \frac{-H_4^2+\rho^4 \Delta \sin^2\theta}{\rho^2 H_3} \Big){\rm d}\phi^2,
\end{equation}
and its determinant is 
$$\bar{g}=\rho^2\sin^2\theta .$$
Here we have set $\rho^2\equiv \sqrt{H_1 H_2}$ and $H_4\equiv B_3H_3$ where $B_3$ is defined in~\eqref{B3}. Next, we introduce the dimensionless parameters ($b,\,c$) such that $p\equiv bm$ and $q\equiv cm$, and other dimensionless parameters defined by $\epsilon^2\equiv Q^2/M^2$, $\mu^2\equiv P^2/M^2$, $\alpha\equiv a/M$ and $x\equiv r/M$. From now on we will adopt ($x,\,M,\,\alpha,\,b,\,c$) as free independent parameters in terms of which the relevant quantities take the following form.
\begin{eqnarray}
	\label{r1}m&=&\frac{4M}{b+c},\qquad \epsilon^2=\frac{4c (c^2-4)}{(b+c)^3},\qquad \mu^2=\frac{4b (b^2-4)}{(b+c)^3},\qquad J=\frac{\sqrt{bc}(bc+4)}{(b+c)^2}M^2\alpha,\\
	\frac{H_1}{M^2}&=& \frac{8(b-2)(c-2)b}{(b+c)^3}+\frac{4(b-2)x}{b+c}+x^2-\frac{2b\sqrt{(b^2-4)(c^2-4)}~\alpha\cos\theta}{(b+c)^2}+\alpha^2\cos^2\theta,\\
	\frac{H_2}{M^2}&=& \frac{8(b-2)(c-2)c}{(b+c)^3}+\frac{4(c-2)x}{b+c}+x^2+\frac{2c\sqrt{(b^2-4)(c^2-4)}~\alpha\cos\theta}{(b+c)^2}+\alpha^2\cos^2\theta,\\
	\label{r1D}\frac{H_3}{M^2}&=& x^2+\alpha^2\cos^2\theta-\frac{8x}{b+c},\qquad \frac{\Delta}{M^2} = x^2+\alpha^2-\frac{8x}{b+c},\\
	\label{r2D}\frac{H_4}{M^3}&=&\frac{2\sqrt{bc}[(bc+4)(b+c)x-4(b-2)(c-2)]\alpha\sin^2\theta}{(b+c)^3}.
\end{eqnarray}
The spacetime admits two horizons namely, $r_\pm=m\pm\sqrt{m^2-a^2},$ obtained by solving $\Delta=0$. This expression is very similar to that of the Kerr BH and it may seem that the event horizon does not depend on the electric and magnetic charges. This is, however, not true. As we explained in the paragraph following Eq.~\eqref{para}, we can express the parameter $m$ in terms of the physical mass $M$, the electric charge $Q$ and the magnetic charge $P$. This shows that the event horizon as well as the radii of the ergo-region depend well on ($M,\,Q,\,P$) even for zero rotation. An expression of $r_+$, which will be denoted by $r_{\text{h}}$, in terms of ($M,\,Q,\,P$) is given in the next subsection in the case where $\epsilon^2\ll 1$ and $\mu^2\ll 1$ [see Eq.~\eqref{h2}].

This applies to all parameters in~\eqref{r1} through~\eqref{r2D}. That is, since $b$ and $c$ may be expressed in terms of $\epsilon^2$ (electric charge) and $\mu^2$ (magnetic charge) on solving the second and third expressions in~\eqref{r1} for $b$ and $c$, the parameters ($m,\,\Delta,\,H_1,\,H_2,\,H_3,\,H_4,\,J$) are all also functions of ($M,\,Q,\,P$). 

Notice that the metric (\ref{met}) has similarity with the rotating Kaluza-Klein solution with dilaton field as discussed in \cite{horo}. The thermodynamic investigations of charged RKKBH reveal interesting results: the temperature of the black hole horizon increases to indefinitely large values as the mass decreases while the entropy of horizon increases with mass too.\\

\subsection*{Physical properties}

In this paper we discuss some physical properties of the projected four dimensional metric~\eqref{met} that have not been discussed in~\cite{npb}, where particularly some thermodynamic entities have been evaluated. The aim of this subsection is identify those useful properties that will allow us to compare~\eqref{met} to well known four dimensional solutions. One obvious property is that the metric~\eqref{met} reduces to the Kerr metric in the case $b=2$ and $c=2$ ($p=2m$ and $q=2m$) where all the charges vanish: $\epsilon^2=0$ and $\mu^2=0$, and, consequently, it reduces to the Schwarzschild metric taking $b=2$, $c=2$ and $\alpha=0$. However, in the absence of magnetic charges ($b=2$), the solution~\eqref{met} never reduces to KNBH. From this point of view, the metric~\eqref{met} is a generalization of the KNBH. The thermodynamical properties of~\eqref{met} have been discussed in~\cite{npb}. 

As we mentioned earlier the parameters $b$ and $c$ may be expressed in terms of $\epsilon^2$ and $\mu^2$ on solving the second and third expressions in~\eqref{r1} for $b$ and $c$. The resulting formulas, expressing $b$ and $c$ as functions of $\epsilon^2+\mu^2$ and $\epsilon^2\mu^2$, are however sizable. First, we set $\eta\equiv b+c$ and $\kappa\equiv bc$. On combining the second and third expressions in~\eqref{r1} we obtain
\begin{equation*}
	(b+c)^2=4~\frac{b^2+c^2-bc-4}{\epsilon^2+\mu^2},
\end{equation*}
which results in
\begin{equation}\label{eta2}
	\eta^2=\frac{4(4+3\kappa)}{4-\epsilon^2-\mu^2}.
\end{equation}
This implies that
\begin{equation}
	\epsilon^2+\mu^2<4,
\end{equation}
no matter the value of the rotation parameter $\alpha$ is. For physical solutions the upper bounds should be $\epsilon^2<1$ and $\mu^2<1$, this shows that unusual solutions with $\epsilon^2\geq 1$ and $\mu^2\geq 1$ might exist in higher dimensional general relativity. However, new larger limits, $\epsilon^2<4$ and $\mu^2<4$ subject to $\epsilon^2+\mu^2<4$, are set. Next, the product of the second and third expressions in~\eqref{r1} yields the cubic equation in $\kappa$
\begin{equation}\label{kappa}
	4 \epsilon ^2 \mu ^2(4+3 \kappa)^3=(4-\epsilon ^2-\mu ^2)^2 \kappa [(4-\epsilon ^2-\mu ^2)\kappa^2 -8
	(2+\epsilon ^2+\mu ^2)\kappa-16 (\epsilon ^2+\mu ^2)],
\end{equation}
where we have used~\eqref{eta2} to eliminate $\eta^2$. Once $\kappa$ is determined from~\eqref{kappa} one obtains an expression for $\eta$ from~\eqref{eta2}. Expressions for $b$ an $c$ are derived upon solving $z^2-\eta~z+\kappa=0$ where $z$ stands for $b$ or $c$.

In the limits $\epsilon^2\ll 1$ and $\mu^2\ll 1$ one can provide corrections to the KNBH of first order in ($\epsilon^2,\,\mu^2$). Of relevant consequences to this work are the event horizon and the outer radius of the ergoregion that are solutions to $\Delta =0$ and $H_3=0$:
\begin{align}
	\label{h}&x_{\text{h}}=\frac{4+\sqrt{16-(b+c)^2\alpha^2}}{b+c},\\
	\label{erg}&x_{\text{erg}}=\frac{4+\sqrt{16-(b+c)^2\alpha^2\cos^2\theta}}{b+c}.
\end{align}
The extremal black hole corresponds to
\begin{equation}\label{ext}
	\frac{16}{(b+c)^2}-\alpha^2=0.
\end{equation}
In the limits $\epsilon^2\ll 1$ and $\mu^2\ll 1$ it is much easier to solve~\eqref{r1} for $b$ and $c$ in terms of ($\epsilon^2,\,\mu^2$), 
\begin{equation}\label{bc}
b\simeq 2+2\mu^2+3\epsilon^2\mu^2,\quad  c\simeq 2+2\epsilon^2+3\epsilon^2\mu^2.
\end{equation}
Finally Eqs.~\eqref{h} and~\eqref{ext} take the forms
\begin{align}
	&x_{\text{h}}\simeq 1+\sqrt{1-\alpha ^2}-\frac{1}{2} \Big(\frac{\sqrt{1-\alpha ^2}+1}{\sqrt{1-\alpha ^2}}\Big) (\epsilon ^2+\mu ^2)+\Big(\frac{2-3 \alpha ^2+2
		(1-\alpha ^2) \sqrt{1-\alpha ^2}}{8 \sqrt{1-\alpha ^2} (1-\alpha ^2)}\Big) (\epsilon ^2+\mu ^2)^2\nonumber\\
	\label{h2}&\qquad-\frac{3}{2} \Big(\frac{\sqrt{1-\alpha ^2}+1}{\sqrt{1-\alpha
			^2}}\Big) \epsilon ^2 \mu ^2,\\
	\label{ext2}&1-\alpha ^2-\epsilon ^2-\mu ^2+\frac{3}{4} (\epsilon ^2-\mu ^2)^2\simeq 0.
\end{align}
We see from~\eqref{ext2} that the first four terms correspond to a doubly charged KNBH. The last term is a correction of second order in ($\epsilon^2,\,\mu^2$). The r.h.s of~\eqref{h2} reduces to the Kerr term if all the charges are zero. We see that even in the limits $\epsilon^2\ll 1$ and $\mu^2\ll 1$ the first three terms of the r.h.s of~\eqref{h2}, which we rewrite as
\begin{equation}
	x_{\text{h}}= 1+\sqrt{1-\alpha ^2}-\frac{1}{2\sqrt{1-\alpha ^2}} (\epsilon ^2+\mu ^2)-\frac{1}{2}~(\epsilon ^2+\mu ^2)+\cdots ,
\end{equation}
provide a correction of first order in ($\epsilon^2,\,\mu^2$) to the value of $x_{\text{h}}$ for the KNBH in the same limits. The correction is the extra term $-(\epsilon ^2+\mu ^2)/2$.\\

\section{General formalism of spin precession in five dimensions\label{secgy}}
In five-dimensional Kaluza-Klein theories the spacetime is equipped with a metric $g_{\alpha\beta}$ independent of the extra spacelike dimension $x^5=\psi$~\cite{KK}
\begin{equation}\label{s1}
{\rm d}s^2=g_{\alpha\beta}(x^{\mu}){\rm d}x^{\alpha}{\rm d}x^{\beta},
\end{equation}
of signature ($+,+,+,-,+$). The known $4+1$ decomposition of the metric~\eqref{s1} leads particularly to the four-dimensional metric $\bar{g}_{\mu\nu}$ in the Einstein frame \cite{KK}
\begin{equation}\label{s2}
\bar{g}_{\mu\nu}=\sqrt{g_{55}}\Big(g_{\mu\nu}-\frac{g_{\mu 5}g_{\nu 5}}{g_{5 5}}\Big),
\end{equation}
where the expression between parentheses is the four-dimensional metric in the Jordan frame. It is worth emphasizing that, as stated in the Introduction, we aim to compare the effects of the gyroscope motion in both the RKKBH and the KNBH, \emph{it is thus imperative to refer to the same frame}. Since the KNBH is expressed in the Einstein frame, it is this frame that we use throughout this work.

The extra dimension $x^5$, being compactified, is unobservable. This implies that any rotation in the Klein circle or any motion in the fifth dimension is also unobservable: The only observable rotation would be that along the spatial coordinates $x^i$. If the stationary metric in endowed with axial symmetry depending only on ($x^1=r,\,x^2=\theta$) and independent of ($x^3=\phi ,\,x^4=t$), the general Killing vector $K^{\alpha }\partial _{\alpha }$ reduces to $K= \partial _t+\Omega \partial_{\phi }$ and its corresponding co-vector (or a 1-form) is given by
\begin{equation}\label{s3}
\bar{K}=\bar{g}_{44}{\rm d}t+\bar{g}_{34}{\rm d}\phi +\Omega (\bar{g}_{34}{\rm d}t+\bar{g}_{33}{\rm d}\phi).
\end{equation}
Consider a test gyroscope attached to an observer moving with four-velocity $u=K/\sqrt{|K^2|}$ along an integral curve of the timelike Killing vector $K$ in a stationary $5$-dimensional spacetime. This is generally not a geodesic motion; in the special case where the motion is geodesic, the precession of the gyroscope is called geodetic precession. The gyroscope is supported by an engine so that it can perform a non-geodesic motion and, for any motion of the gyroscope, $\Omega$ has to be held constant for $K$ to be a Killing vector. The spin of the gyroscope can be represented by the vorticity field of the Killing
congruence. As shown in the appendix section, the general spin precession one-form $\bar{\Omega}_{p}$ of the test gyroscope is given by \cite{Str}
\begin{equation}\label{ofp}
\bar{\Omega}_{p}=\frac{1}{2K^{2}}\ast ( \bar{K}\wedge {\rm d}\bar{K}),
\end{equation}
where $\ast $ represents the Hodge star operator and $\wedge $ is the wedge product. Note that the quantity $\ast ( \bar{K}\wedge {\rm d}\bar{K})$ can be regarded as a measure of the {\textquotedblleft absolute\textquotedblright} rotation. The gyroscope is moving in five dimensions and we are considering the projection of this motion onto the four dimensional spacetime.

Using~\eqref{ofp}, we can first evaluate the one-form of the precession frequency $\bar{\Omega}_p$ then its vector $\vec{\Omega}_p$, representing the overall rotation in the four-dimensional spacetime, by
\begin{equation}\label{s4}
\vec{\Omega}_p=\frac{\pm \epsilon_{ab}}{2\sqrt{|\bar{g}|}\big(\bar{g}_{44}+2\Omega\bar{g}_{34}+\Omega^2\bar{g}_{33}\big)}\Big[\bar{g}_{44}\bar{g}_{34,a}-\bar{g}_{34}~\bar{g}_{44,a}+\Omega\Big(\bar{g}_{44}\bar{g}_{33,a}-\bar{g}_{33}~\bar{g}_{44,a}\Big)+\Omega^2\Big(\bar{g}_{34}~\bar{g}_{33,a}-\bar{g}_{34}~\bar{g}_{34,a}\Big)\Big]\partial_{b},
\end{equation}
where $\epsilon_{ab}$ is the totally antisymmetric symbol. The overall sign $\pm$ is due to the different conventions in the definition of the Hodge star\footnote{A definition of the Hodge star is
\begin{equation*}
	\ast ({\rm d}x^{I_1}\wedge \cdots \wedge {\rm d}x^{I_p})=\frac{\sqrt{|g|}}{(n-p)!}~\epsilon
	_{\nu_1\cdots\nu_{n-p}\mu_1\cdots\mu_{p}}g^{\mu_1I_1}\cdots g^{\mu_{p}I_p}{\rm d}x^{\nu_1}\wedge \cdots \wedge  x^{\nu_{n-p}}.
\end{equation*}
\label{foot1}} and the definitions $\epsilon_{0123}=+1$ and $\epsilon_{1234}=+1$ as we are labeling the time coordinate by $x^4$ instead ot $x^0$. In the limit, $\Omega=0$, one obtains the expression of Lense-Thirring precession frequency in five dimensions.\\

\section{Spin precession of a test gyroscope in RKKBH}

\begin{figure*}[h]
	\centering
	\includegraphics[width=0.43\textwidth]{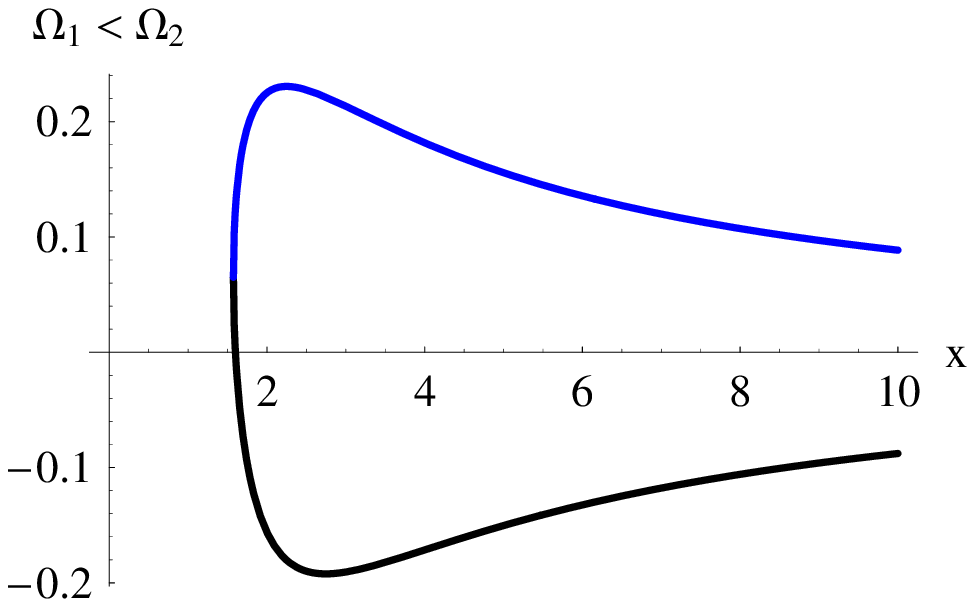}
	\includegraphics[width=0.43\textwidth]{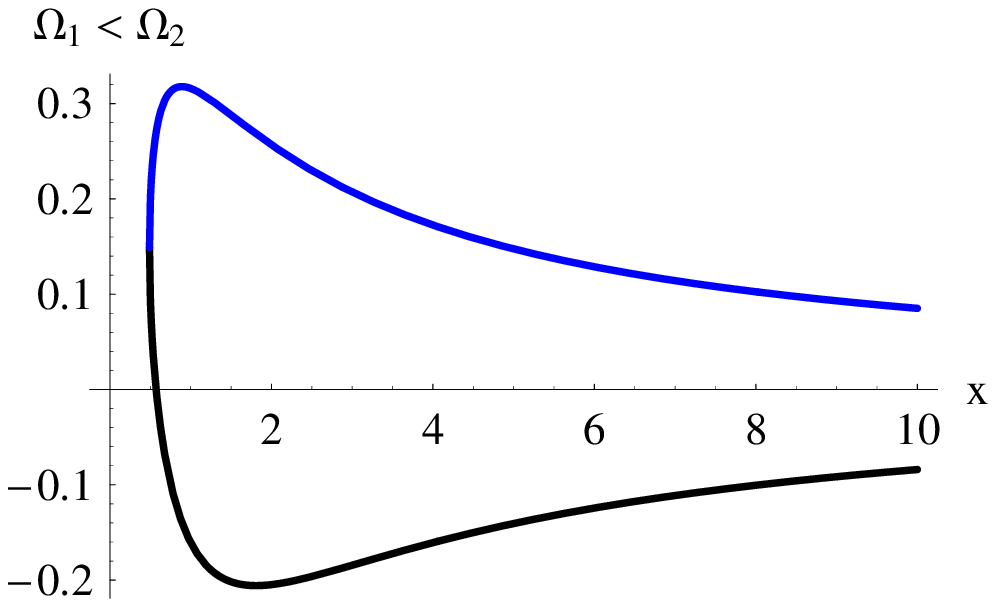} \\
	\caption{\footnotesize{Plots of ($\Omega_1,\,\Omega_2$)~\eqref{mM} in the units of $1/M$, that is, plots of the dimensionless entities ($M\Omega_1,\,M\Omega_2$) versus $x=r/M$ for $\theta=\pi/2$ and $\alpha=1/5$. In the left panel we took $b=2\;\&\;c=3$ (the analogous to KNBH with $\epsilon^2=12/25$ and $\mu^2=0$) and in the right panel we took $b=c=7$ ($\epsilon^2=\mu^2=45/98$). The two curves meet at $x=x_{\text{h}}$.}}\label{Fig1}
\end{figure*}

We consider the following quantities:
\begin{equation}\label{s4-1}
\Omega_\theta\equiv\bar{g}_{44}\bar{g}_{34,r}-\bar{g}_{34}~\bar{g}_{44,r}+\Omega\Big(\bar{g}_{44}\bar{g}_{33,r}-\bar{g}_{33}~\bar{g}_{44,r}\Big)+\Omega^2\Big(\bar{g}_{34}~\bar{g}_{33,r}-\bar{g}_{34}~\bar{g}_{34,r}\Big),
\end{equation}
\begin{equation}\label{s4-2}
\Omega_r\equiv\bar{g}_{44}\bar{g}_{34,\theta}-\bar{g}_{34}~\bar{g}_{44,\theta}+\Omega\Big(\bar{g}_{44}\bar{g}_{33,\theta}-\bar{g}_{33}~\bar{g}_{44,\theta}\Big)+\Omega^2\Big(\bar{g}_{34}~\bar{g}_{33,\theta}-\bar{g}_{34}~\bar{g}_{34,\theta}\Big),
\end{equation}
where $\Omega$ is constant bounded by the constraint that $K= \partial _t+\Omega \partial_{\phi }$ is timelike, that is,
\begin{equation}
\bar{g}_{44}+2\Omega\bar{g}_{34}+\Omega^2\bar{g}_{33}<0,
\end{equation}
resulting in
\begin{equation}\label{const}
\min(\Omega_1(r,\theta))<\Omega<\max(\Omega_2(r,\theta)).
\end{equation}
Thus, $\Omega$ is any number smaller than the maximum value of the function $\Omega_2(r,\theta)$ and bigger than the minimum value of the function $\Omega_1(r,\theta)$ where
\begin{equation}\label{mM}
\Omega_1=\frac{H_3}{-\rho^2\sqrt{\Delta}\sin\theta -H_4},\qquad \Omega_2=\frac{H_3}{\rho^2\sqrt{\Delta}\sin\theta -H_4}.
\end{equation}
We call these two functions, which are depicted in Fig.~\ref{Fig1}, the limit frequencies for timelike motion. Since on the horizon we have $\Delta=0$, this results in $\Omega_1=\Omega_2=-H_3/H_4$ at $x=x_{\text{h}}$.

We intend to investigate the behavior of the norm of the vector $\vec{\Omega}_p$,
\begin{equation}\label{norm}
|\vec{\Omega}_p|=\frac{\sqrt{\bar{g}_{11}\Omega_r^2+\bar{g}_{22}\Omega_\theta^2}}{2\sqrt{|\bar{g}|}~\big|\bar{g}_{44}+2\Omega\bar{g}_{34}+\Omega^2\bar{g}_{33}\big|},
\end{equation}
where the presence of the metric coefficients ($\bar{g}_{11}=\rho^2/\Delta,\,\bar{g}_{22}=\rho^2$) is to take into account the fact that ($\partial_r,\,\partial_{\theta}$)~\eqref{s4} are not unit vectors.
For the metric~\eqref{met}, ($\Omega_\theta,\,\Omega_r$) are given by
\begin{eqnarray}\label{s4-3}
\Omega_\theta &=&\frac{1}{H_3^2\rho^4}\Big\{2H_3^2H_4(r  -m) -
\frac{a\sqrt{bc}(4+bc)H_3^3m\sin^2\theta}{2(b+c)}+\Omega\Big[4rH_3H_4^2 -4mH_3H_4^2 -\frac{a\sqrt{bc}(4+bc)H_3^3 H_4m\sin^2\theta}{b+c} \nonumber\\&&   +2H_3^2(r-m)\rho^4\sin^2\theta+
H_3^2[H_2((b-2)m+2r)+H_1((-2+c)m+2r)]\Delta \sin^2\theta+4mH_3\rho^4\Delta\sin^2\theta
-4rH_3\rho^4\Delta\sin^2\theta  \Big]\nonumber\\&&+\Omega^2\Big[H_3H_4\rho^2\Big(  2(r-m)\rho^2+\frac{[H_2((b-2)m+2r)+H_1((c-2)m+2r)]\Delta}{\rho^2}   \Big)\sin^2\theta - \frac{a\sqrt{bc}(4+bc)H_3H_4^2m\sin^2\theta}{2(b+c)}\nonumber\\&&-\frac{a\sqrt{bc}(4+bc)mH_3\rho^4\sin^2\theta}{2(b+c)} +2H_4(r-m)(H_4^2-\rho^4\Delta\sin^2\theta) \Big]   \Big\},
\end{eqnarray}
\begin{eqnarray}\label{s4-4}
\Omega_r &=& \frac{1}{H_3^2\rho^4}\Big\{
\frac{a\sqrt{bc}H_3^3m[(b-2)(c-2)m-(4+bc)r]\cos\theta\sin\theta}{b+c}-2a^2H_3^2H_4\cos\theta\sin\theta
\nonumber\\&&+\Omega\Big[
\frac{2a\sqrt{bc}H_3^2H_4m[(b-2)(c-2)m-(4+bc)r]\cos\theta\sin\theta}{b+c}-2a^2H_3H_4^2\cos\theta\sin\theta +4a^2H_3\rho^4\Delta\cos\theta\sin^3\theta\nonumber\\&& \frac{aH_3^2\Delta\big[H_1[\sqrt{b^2-4}c\sqrt{c^2-4}m+4a(b+c)\cos\theta]+H_2[-b\sqrt{(b^2-4)(c^2-4)}m+4a(b+c)\cos\theta]\big]\sin^3\theta}{2(b+c)} \nonumber\\&&+H_3^2\rho^4\Delta \sin(2\theta)  \Big]
+\Omega^2\Big[  \frac{a\sqrt{bc}H_3H_4^2m[(b-2)(c-2)m-(4+bc)r]\cos\theta\sin\theta}{b+c}\\&&+
\frac{a\sqrt{bc}H_3m((b-2)(c-2)m-(4+bc)r)\rho^2\cos\theta\sin^3\theta}{b+c}
-2a^2H_4\cos\theta\sin\theta(H_4^2-\rho^4\Delta\sin^2\theta) +2H_3H_4\rho^2\Delta\sin\theta\nonumber\\&&\times
\Big(   \rho^2\cos\theta-\frac{a[H_1(\sqrt{(b^2-4)(c^2-4)}mc+4a(b+c)\cos\theta)+H_2(-b\sqrt{(b^2-4)(c^2-4)}m+4a(b+c)\cos\theta)]\sin^2\theta}{4(b+c)\rho^2}  \Big)\Big]     \Big\},\nonumber
\end{eqnarray}
while the expression in the denominator of Eq. (\ref{norm}), $2\sqrt{|\bar{g}|}~\big|\bar{g}_{44}+2\Omega\bar{g}_{34}+\Omega^2\bar{g}_{33}\big|$, simplifies to
\begin{equation}\label{denom}
\frac{2\big|H_3^2+2\Omega H_3 H_4+\Omega^2(H_4^2-\rho^4\sin^2\theta\Delta)\big|\sin\theta}{|H_3|}.
\end{equation}
For $\Omega$ held constant, its zeros will be denoted by $x_1$ and $x_2$:
\begin{align}\label{x1x2}
&x_{\text{h}}<x_1<x_{\text{erg}}<x_2\quad\text{if}\quad\Omega\neq\omega_h,\\
&x_{\text{h}}=x_1<x_{\text{erg}}<x_2\quad\text{if}\quad\Omega = \omega_h\nonumber,
\end{align}
where $\omega_h\equiv\omega(x=x_h)$ and $\omega(x)\equiv -\bar{g}_{34}/\bar{g}_{33}$ is the ZAMO's angular velocity satisfying
\begin{multline}\label{ZAMO}
\omega=-\frac{H_3H_4}{H_4^2-\rho^4\Delta\sin^2\theta},\qquad \Omega_{1}\Omega_{2}=-\frac{H_3\omega}{H_4},\\ M\omega_h=-M\frac{H_3}{H_4}\Big|_{x=x_h}=\frac{(b+c)^3\alpha}{2\sqrt{bc}[(bc+4)(b+c)x_h-4(b-2)(c-2)]}.
\end{multline}

\begin{figure*}[h]
	\centering
	\includegraphics[width=0.43\textwidth]{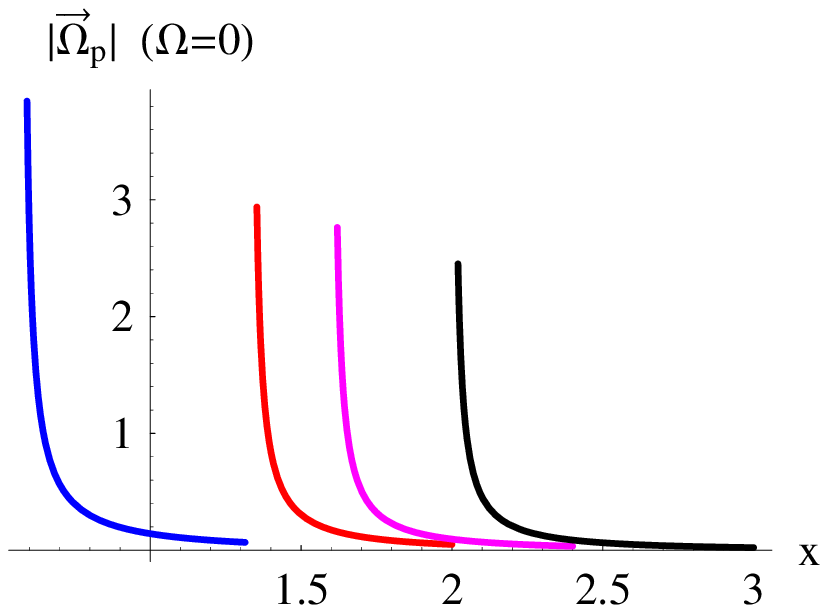}
	\includegraphics[width=0.43\textwidth]{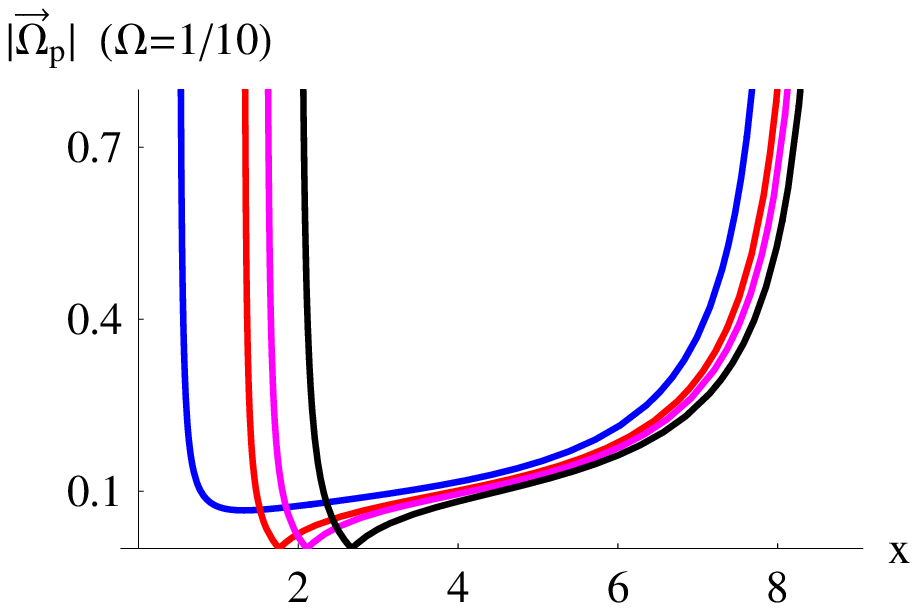} \\
	\caption{\footnotesize{Plots of $|\vec{\Omega}_p|$~\eqref{norm} in the units of $1/M$, that is, plots of the dimensionless norm $M|\vec{\Omega}_p|$ versus $x=r/M$ for $\theta=\pi/2$ and $\alpha=1/5$. We took $b=c=7$ ($\epsilon^2=\mu^2=45/98$) for the blue plot, $b=c=3$ for the red plot ($\epsilon^2=\mu^2=5/18$), $b=2\;\&\;c=3$ for the magenta plot (the analogous to KNBH with $\epsilon^2=12/25$ and $\mu^2=0$), and $b=c=2$ for the black plot corresponding to the Kerr black hole. In the left panel we took $\Omega=0$. The norm $|\vec{\Omega}_p|$ diverges on the surface of the ergoregion $x=x_{\text{erg}}$ and the gyroscope may remain on a timelike curve for all $x>x_{\text{erg}}$. As the black hole becomes more and more charged, the three-space outside the ergoregion extends. In the right panel we took $\Omega=1/10$. The norm $|\vec{\Omega}_p|$ diverges at the two zeros $x_1$ and $x_2$~\eqref{x1x2} of the denominator of~\eqref{norm}, given in~\eqref{denom}, and the gyroscope may remain on a timelike curve only for $x$ taken between these zeros. As the black hole becomes more and more charged, both zeros decrease and the three-space between them extends.}}\label{Fig2}
\end{figure*}

A series expansion of $|\vec{\Omega}_p|$ for $\alpha^2\ll 1$, $\epsilon^2\ll 1$, $\mu^2\ll 1$ and $\theta=\pi/2$ yields
\begin{align}
|\vec{\Omega}_p|=&\bigg|\frac{\Omega(r-3 M) }{r-2 M-r^3 \Omega ^2}\bigg|~
 \bigg\{1+\frac{M^2 \left(3 r^2 \Omega
		^2-1\right) \left(2 M-r-r^3 \Omega ^2\right)}{r^2(3 M-r) \left(2 M-r+r^3 \Omega ^2\right) \Omega }~ \alpha\nonumber\\
	\qquad & +\frac{M^2 \left[12 M^3 \Omega ^2 r+\Omega
		^2 \left(r^2 \Omega ^2-1\right) r^4+M \left(2-r^2 \Omega ^2+r^4 \Omega ^4\right) r-2 M^2 \left(2-2 r^2 \Omega ^2+9 r^4 \Omega ^4\right)\right]}{r
		(3 M-r) \left(2 M-r+r^3 \Omega ^2\right)^2}~\alpha ^2\nonumber\\
\label{norms1}	\qquad &-\frac{M (2 M^2-3 M r+13 M \Omega ^2 r^3-2 \Omega ^2 r^4)}{2 r (3 M-r) (2 M-r+r^3
		\Omega ^2)}~ (\epsilon ^2+\mu ^2)+\cdots\bigg\},\qquad \Omega\neq 0,\\
\label{norms2}|\vec{\Omega}_p|=&\bigg|\frac{M^2 \alpha}{r^2 (r-2 M)}\bigg|~ \bigg\{1+\frac{M (6 M-5 r)}{2 r (r-2 M)}~ (\epsilon
^2+\mu ^2)+\cdots\bigg\},\qquad \Omega = 0.
\end{align}
These expressions have been derived using~\eqref{bc}. In the case $\Omega\neq 0$, even if the BH is not rotating ($\alpha =0$), there is a nonvanishing contribution to $|\vec{\Omega}_p|$ as we see from~\eqref{norms1}, which for the Schwarzschild BH reduces to the factor
\begin{equation}\label{fac}
|\vec{\Omega}_p|=\bigg|\frac{\Omega(r-3 M) }{r-2 M-r^3 \Omega ^2}\bigg|,
\end{equation}
in~\eqref{norms1}. In the Schwarzschild spacetime it is known that \emph{if the gyroscope moves along a circular geodesic} then its angular velocity, or Kepler frequency $\Omega=\Omega_{\text{Kep}}$, is related to the radius of the circle  by $\Omega\equiv\Omega_{\text{Kep}}=\sqrt{M/r^3}$. On replacing $\Omega$ by $\sqrt{M/r^3}$ in~\eqref{fac} we obtain $|\vec\Omega_p|=\Omega_{\text{Kep}}=\sqrt{M/r^3}$, that is, the  precession frequency will be the same as the Kepler frequency. If in the Schwarzschild spacetime the gyroscope, supported by an engine, rotates with an angular velocity $\Omega\neq\Omega_{\text{Kep}}$, then $|\vec\Omega_p|\neq\Omega$. No, if the gyroscope has no angular velocity in the stationary spacetime, $\Omega =0$ ($K=\partial_t$), while the main contribution to $|\vec{\Omega}_p|$ comes from rotation, as we see from~\eqref{norms2}, there are contributions from the electric ($\epsilon^2$) and magnetic ($\mu^2$) charges as well.  

Now back to the general expression~\eqref{norm}. For $\Omega=0$, there is nothing special as Fig.~\ref{Fig2} reveals: The gyroscope may remain on a timelike curve for all $x>x_{\text{erg}}$. As the black hole becomes more and more charged, the three-space outside the ergoregion extends. For $\Omega\neq 0$, as shown in the right panel of Fig.~\ref{Fig2} where $\Omega =1/10$, the norm $|\vec{\Omega}_p|$ diverges at the two zeros $x_1$ and $x_2$~\eqref{x1x2} of the denominator of~\eqref{norm}, given in~\eqref{denom}, and, for $\Omega$ held constant, the gyroscope may remain on a timelike curve only for $x$ taken between these zeros. As the black hole becomes more and more charged, both zeros decrease and the three-space between them extends.

Notice the presence of a point $x_{\text{min}}$ where $|\vec{\Omega}_p(x_{\text{min}})|=0$, that is, a point where $\Omega_\theta(x_{\text{min}})=0$ and $\Omega_r(x_{\text{min}})=0$. Such a point may offer a way for distinguishing KNBH and RKKBH. Another way to distinguish these BHs is to consider the minimum value of $M\Omega_1$ and the maximum value of $M\Omega_2$~\eqref{mM} versus $\epsilon^2$, as depicted in Fig~\ref{Fig1} and subsequent figures. \\

\section{Distinguishing Kerr-Newman and Rotating Kaluza-Klein black holes}

\subsection{$|\vec{\Omega}_p(x_{\text{min}})|=0$}

The metric of the KNBH may be brought to the form~\eqref{met} with
\begin{align}
&\frac{\rho_{\text{KN}}^2}{M^2}= x^2+\alpha^2\cos^2\theta, & & \frac{\Delta_{\text{KN}}}{M^2} = x^2-2x+\epsilon^2+\alpha^2,\\
&\frac{H_{3\,\text{(KN)}}}{M^2}= x^2+\alpha^2\cos^2\theta-2x+\epsilon^2, & & \frac{H_{4\,\text{(KN)}}}{M^3}=\alpha (2x-\epsilon^2)\sin^2\theta.
\end{align}

\begin{figure*}[h]
	\centering
	\includegraphics[width=0.40\textwidth]{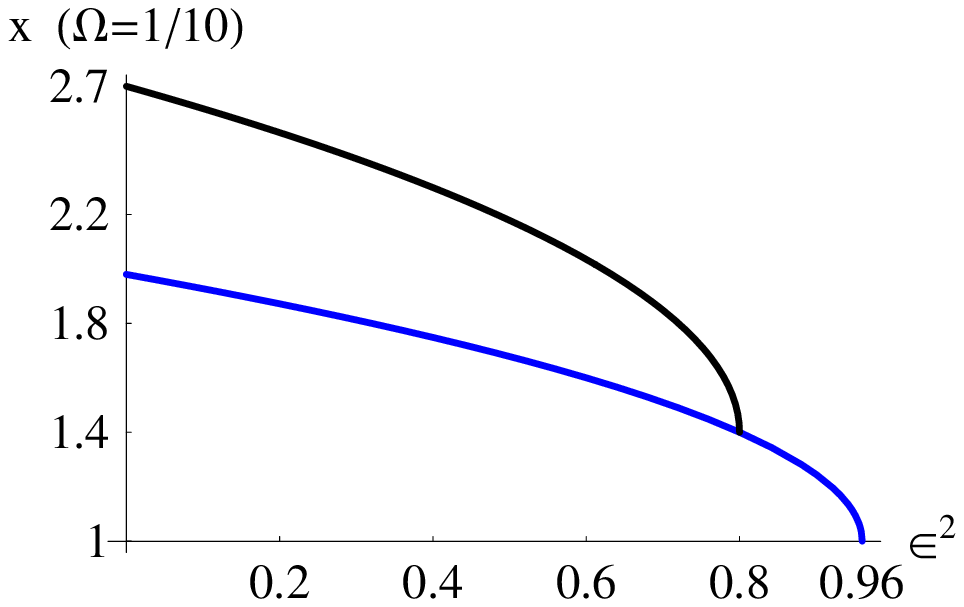}
	\includegraphics[width=0.40\textwidth]{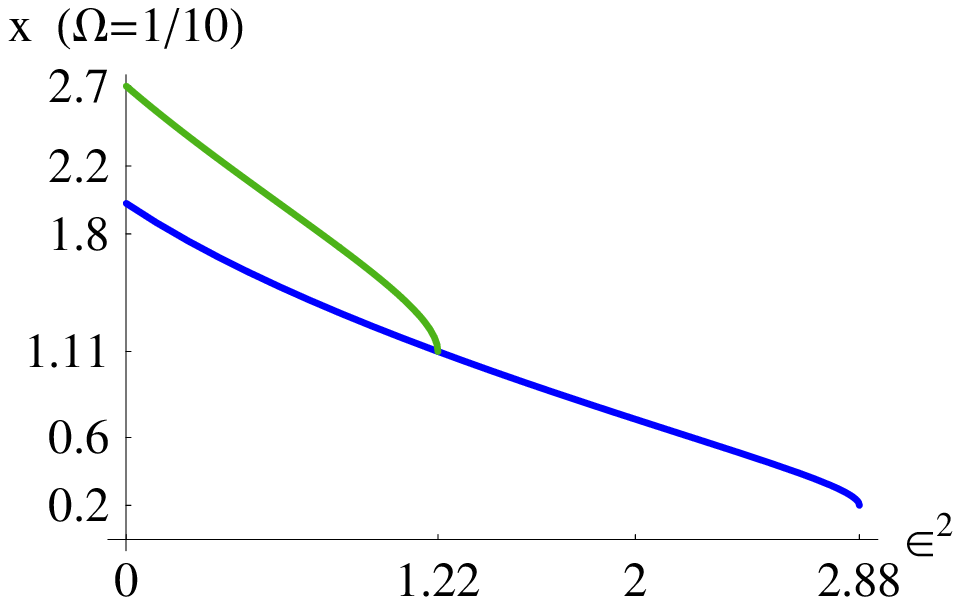}\\
	\includegraphics[width=0.40\textwidth]{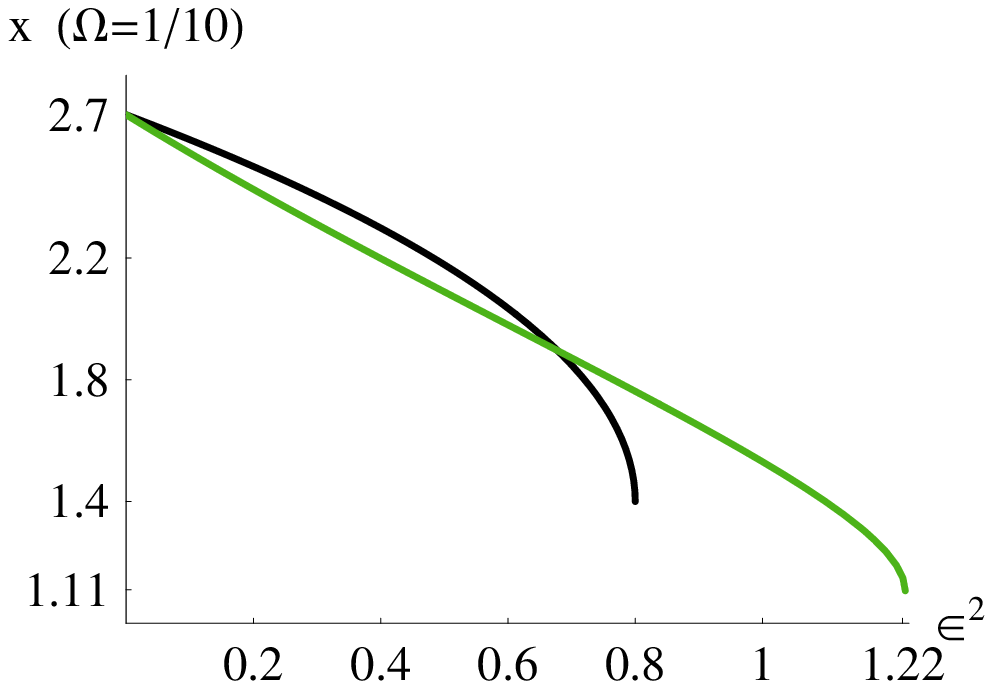}
	\includegraphics[width=0.40\textwidth]{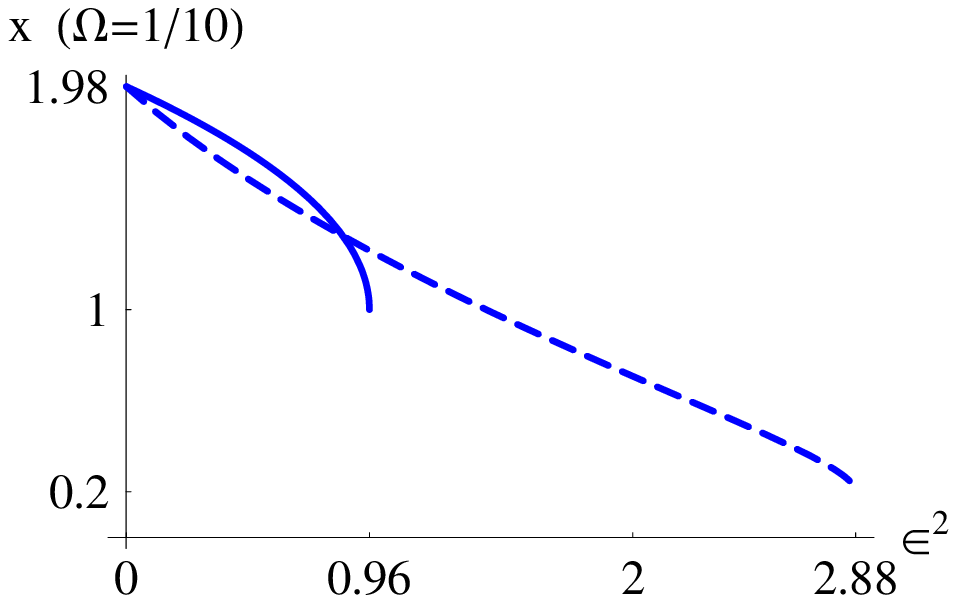} \\
	\caption{\footnotesize{Plots of $x_{\text{min}}$ such that $|\vec{\Omega}_p(x_{\text{min}})|=0$, the event horizon $x_{\text{h}}$ and the outer radius of the ergo-region versus $\epsilon^2$ for $\theta=\pi/2$, $\alpha=1/5$ and $\Omega=1/10$. Upper Left Panel: The black plot depicts $x_{\text{min}}(\epsilon^2)$ and the blue plot depicts $x_{\text{h}}(\epsilon^2)$ for the KNBH. The blue plot ends at the point ($0.96,\,1$) corresponding to the extremal KNBH. Upper Right Panel: The green plot depicts $x_{\text{min}}(\epsilon^2)$ and the blue plot depicts $x_{\text{h}}(\epsilon^2)$ for the RKKBH with no magnetic charge ($\mu^2=0$). The blue plot ends at the point ($2.88,\,0.2$) corresponding to the extremal RKKBH. Lower Left Panel: The black plot depicts $x_{\text{min}}(\epsilon^2)$ for the KNBH and the green plot depicts $x_{\text{min}}(\epsilon^2)$ for the RKKBH. These are the same plots of the Upper Left and Upper Right Panels combined with the horizon plots deleted. In both the Upper Left and Upper Right Panels the curve $x_{\text{min}}(\epsilon^2)$ meets the $x$-axis at 2.7 corresponding to Kerr BH. Lower Right Panel: The continuous plot depicts the outer radius of the ergo-region for the KNBH and the dashed plot depicts the outer radius of the ergo-region for the RKKBH. These are the same plots of the Upper Left and Upper Right Panels combined with the $x_{\text{min}}(\epsilon^2)$ plots deleted.}}\label{Fig3}
\end{figure*}

We consider a KNBH and a RKKBH with no magnetic charge ($\mu^2=0$). In Fig.~\ref{Fig3} we depict the graphs of $x_{\text{min}}(\epsilon^2)$ where $|\vec{\Omega}_p(x_{\text{min}})|=0$. The graphs of the event horizon $x_{\text{h}}$ versus $\epsilon^2$ are also shown. We are interested in the region outside the event horizon. For the numerical set used in Fig.~\ref{Fig3}, the values of $x_{\text{min}}$ range from 2 to 2.7, the range of the electric charge where $x_{\text{min}}\geq x_{\text{h}}$ is, however, much larger for a RKKBH.

We do not expect the charge of a black hole to exceed its mass, we focus on the physical region corresponding to $\epsilon^2\ll 1$. In this case, as we see from the right panel of Fig.~\ref{Fig3}, a moving gyroscope, following a time like path with an angular velocity $\Omega\neq 0$, may reveal the nature of the BH as follows. To be more precise, we provide the calculations for $\epsilon^2=1/100$. This yields $x_{\text{min}}=2.66275$ for a KNBH and $x_{\text{min}}=2.65781$ for a RKKBH, which do not depend on the mass of the BH and correspond to $\Delta x=0.00493387$. Introducing the relevant physical constants we obtain
\begin{equation}\label{deltar}
\Delta r=\frac{GM}{c^2}~\Delta x,
\end{equation}
where $G=6.673\times 10^{-11}$ and $c=299792458$ in SI units. For a BH with one solar mass ($M_\odot=1.9888\times 10^{30}$ kg), $\Delta r=7.3$ m and for a BH with one million solar masses $\Delta r=7.3\times 10^{6}$ m. In terms of $r$, the gyroscope will detect no spin precession, corresponding to a vanishing value of $|\vec{\Omega}_p|$, at $r_{\text{min}}=3.93188\times 10^{9}$ m if it is moving along a timelike path in a KNBH, if, otherwise, $|\vec{\Omega}_p|$ vanishes at some smaller value of $r$, such that $\Delta r=7.3\times 10^{6}$ m, this should correspond to a RKKBH with no magnetic charge.
\begin{figure*}[h]
	\centering
	\includegraphics[width=0.43\textwidth]{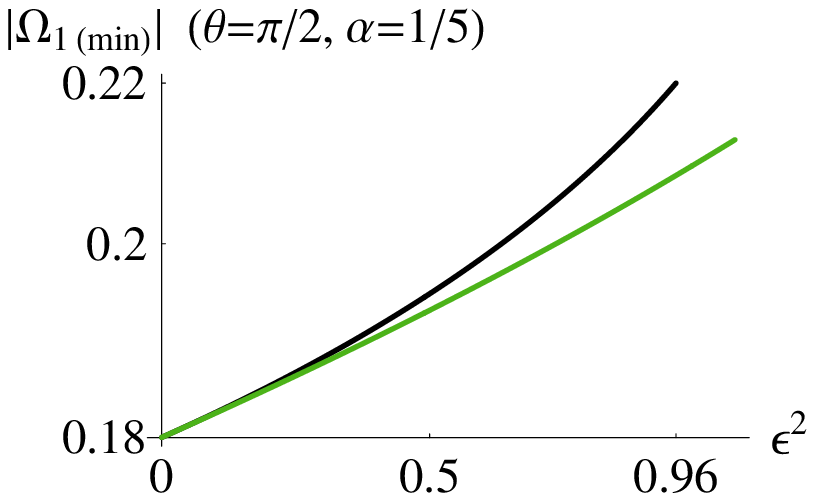}
	\includegraphics[width=0.43\textwidth]{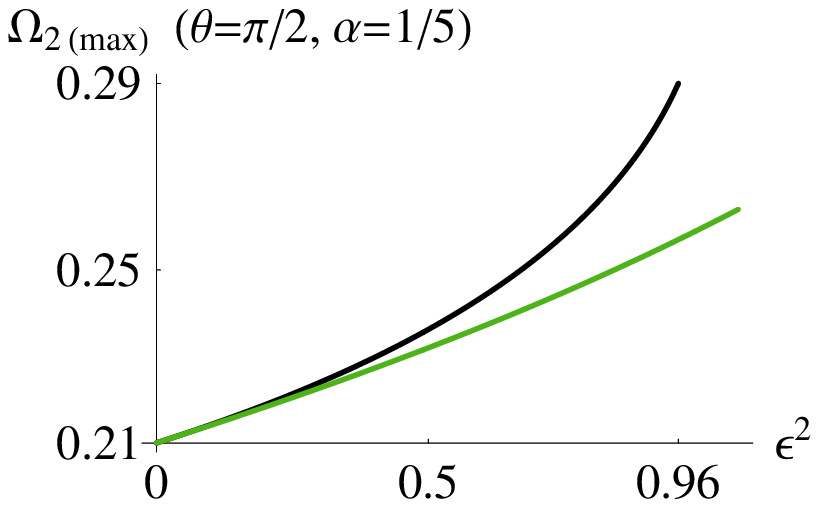} \\
	\caption{\footnotesize{Left Panel: Plot of the absolute value the minimum of $\Omega_1$~\eqref{mM} in the units of $1/M$ (plot of $M|\Omega_{1\text{(min)}}|$) versus $\epsilon^2=Q^2/M^2$ for $\theta=\pi/2$ and $\alpha=1/5$. The black plot represents a KNBH and the green plot represents a RKKBH. Right Panel: Plot of the maximum of $\Omega_2$~\eqref{mM} in the units of $1/M$ (plot of $M\Omega_{2\text{(max)}}$) versus $\epsilon^2$ for $\theta=\pi/2$ and $\alpha=1/5$. The black plot represents a KNBH and the green plot represents a RKKBH with $\mu^2=0$. The green plots extend up to $\epsilon^2=2.88$, which is the value of $\epsilon^2$ for an extremal RKKBH (Fig.~\ref{Fig3}).}}\label{Fig4}
\end{figure*}

\subsection{$M\min(\Omega_1(r,\theta))$ and $M\max(\Omega_2(r,\theta))$}
Another way to be able to distinguish a KNBH and a RKKBH is to compare the extrema of the dimensionless functions ($M\Omega_{1},\,M\Omega_{2}$) for both BHs. In Fig.~\ref{Fig4} we depict the maximum values of ($M|\Omega_{1}|,\,M\Omega_{2}$) versus $\epsilon^2=Q^2/M^2$. For $\epsilon^2=1/100$, we have $M\Omega_{2\text{(max)}}=0.2094775$ for KNBH and $M\Omega_{2\text{(max)}}=0.2094763$ for a RKKBH with $\mu^2=0$. These vales, which are independent of the mass $M$ of the BH, show that for $0.2094763<M\Omega\leq 0.2094775$ a gyroscope in the geometry of a KNBH can still follow a prograde timelike path while this is not possible in the geometry of a RKKBH. For the same value of $\epsilon^2$ we obtain $M\Omega_{1\text{(min)}}=-0.1792895775\simeq -0.1792896$ for KNBH and $M\Omega_{1\text{(min)}}=-0.1792890227\simeq -0.1792890$ for a RKKBH with $\mu^2=0$. This shows that for $-0.1792896\leq M\Omega< -0.1792890$ a gyroscope in the geometry of a KNBH can still follow a retrograde timelike path while this is not possible in the geometry of a RKKBH.

We have chosen $\epsilon^2=1/100$ relatively small because we believe most BHs are lightly charged. For this value of $\epsilon^2$ we see that $M\min(\Omega_1(r,\theta))$ and $M\max(\Omega_2(r,\theta))$ may differ only in the seventh decimal. From an experimental point of view, it may reveal hard, but not possible, to perform such an experiment. However, as it is clear from Fig.~(4), had we chosen a higher value of $\epsilon^2$, $M\min(\Omega_1(r,\theta))$ and $M\max(\Omega_2(r,\theta))$ would differ in a decimal order lower than 7.\\

\section{Discussion}

In this paper, we have extended the analysis of gyroscope precession frequency to five dimensional charged rotating black holes in Kaluza-Klein theory. This phenomenon is related with the stationary gyroscopes moving along timelike curves in a stationary black hole spacetimes. First we derived the general precession frequency formula for test gyroscopes valid for general five dimensional rotating black holes, by dimensionally reduction to four dimensions. From empirical perspective, we studied the magnitude of the precession frequency vector associated with test gyroscopes in KK spacetime, $|\vec{\Omega}_p |$, and the limit frequencies for timelike motion, ($\Omega_{1},\,\Omega_{2}$). We have shown that $|\vec{\Omega}_p |$ may vanish if $\Omega\neq 0$ and that this fact can be used to distinguish astrophysical black holes. We have also shown how the extreme values of ($\Omega_{1},\,\Omega_{2}$) may help distinguishing astrophysical black holes. Both these schemes are mass-independent.

There are few important points of note: the $|\vec{\Omega}_p |$  diverges at two spatial locations outside the event horizon, enclosing the outer radius of the ergoregion. However if $\Omega=0$, than the norm $|\vec{\Omega}_p |$ diverges at a single location only, which is the outer radius of the ergoregion. Moreover, the angular speed $\Omega$ of the stationary gryoscopes takes both positive and negative values, depicting the gyroscopes moving around the black hole in prograde and retrograde orbits respectively. The maximum of $\Omega_2$ occurs much closer to the horizon as compared to the minimum of $\Omega_1$. Ultimately, as the gyroscope approaches the horizon, both  $\Omega_1$ and  $\Omega_2$ approach the ZAMO's angular velocity.

The extra dimension is an important concept in modern gravity theory and there is no experiment to prove or disprove the hypothesis. In our work, we found that a rotating KK black hole is always different from a Kerr-Newman black hole, this implies that we can check the hypothesis of extra dimension by the analysis of the gyroscope precession frequency. Recall that the hypothesis of KK asserts that the extra dimension is compactified throughout the whole spacetime and, consequently, the $4+1$ decomposition remains valid for the whole range of coordinates. The construction of KKBHs is entirely based on this hypothesis. This implicitly assumes that the gravitational field, particularly near and outside the outer radius of the ergoregion, is not high enough to allow for probing the extra dimension.

On the other hand, it is considered that graviton can play an important role to investigate extra dimensions, so gravitational perturbation could include some critical information from the property of spacetime with extra dimensions. We will work on the gravitational perturbation effect on the gyroscope precession in a subsequent work.\\

\section*{Appendix: Precession frequency\label{secaa}}
\renewcommand{\theequation}{A.\arabic{equation}}
\setcounter{equation}{0}
The four-velocity of an observer at rest along an integral curve $\gamma$ of $K$ is
\begin{equation}\label{a1}
u=\frac{K}{\sqrt{|K^2|}}.
\end{equation}
Let $e_4=u$ and $e_i$ ($i$: $1\to 3$) form an orthonormal tetrad: $<e_\mu,e_\nu>=\eta_{\mu\nu}$ [($\mu,\nu$): $1\to 4$] where $\eta_{\mu\nu}={\rm diag}(-1,1,1,1)$ and $<,>$ denotes scalar product. As time evolves we want that the three elements of the triad $e_i$ remain perpendicular to each  other and each of which remains perpendicular to $u\propto K$. The only transport machinery along $\gamma$ that preserves orthogonality is the Lie derivative. Thus, we choose the triad $e_i$ such that
$L_K e_i=0$ and along with $L_K e_{4}\propto L_K K\equiv 0$, which is identically satisfied, we can write  
\begin{equation}\label{key}
L_K e_{\mu}=0.
\end{equation}
This realizes what is called a Copernican system~\cite{Str}. Said otherwise, the basis vectors $e_{i}$ are tied to an inertial system far from the source (BH) which is fixed relative to the distant stars~\cite{Grav}.

The spin $S$ of the gyroscope obeys the equation~\cite{Str,Grav}
\begin{equation}\label{a2}
\nabla_u S=<S,\nabla_u u>u\;\text{ with }\;<S,u>=0,
\end{equation}
where $\nabla_u u$ is the acceleration of the gyroscope moving along an integral curve $\gamma$ of $K$. This acceleration is generally nonzero. By $<S,u>=0$, $S$ is a purely spatial vector $S^4=0$ and $S=S^ie_i$. Now, evaluating ${\rm d}S^i/{\rm d}\tau =\nabla_u <e^i,S>$ where $e^{\mu}=\eta^{\mu\nu}e_{\nu}$ ($e^{i}=e_{i}$), results in
\begin{equation}\label{a3}
\frac{{\rm d}S^i}{{\rm d}\tau}=S^j <\nabla_u e_i,e_j>=S^j \omega_{ij}\quad\text{with}\quad\omega_{ij}:=<\nabla_u e_i,e_j>.
\end{equation}
Here we have dropped the term $<e_i,\nabla_u S>$ which is zero by~\eqref{a2}. Now, since $<e_i,e_j>=\eta_{ij}$ we have that $\omega_{ij}$ is anti-symmetric: $\omega_{ij}=-\omega_{ji}$. Thus the right-hand side of~\eqref{a3} can be put in the form $\varepsilon_{ijk}S^j\Omega^k$ where $\vec{\Omega}_p=\Omega^k e_k$ is the spin's angular velocity of precession relative to the Copernican system we defined above. This is related to $\omega_{ij}$ by
\begin{equation}\label{a4}
\omega_{ij}=\varepsilon_{ijk}\Omega^k,
\end{equation}
where $\varepsilon_{ijk}$ is the totally anti-symmetric symbol. This yields using the property $\varepsilon^{ijl}\varepsilon_{ijk}=2\delta^l_k$
\begin{equation}\label{a5}
\Omega^l=\frac{\varepsilon^{ijl}\omega_{ij}}{2}=\frac{\varepsilon^{ijl}<\nabla_u e_i,e_j>}{2}=\frac{\varepsilon^{ijl}<\nabla_K e_i,e_j>}{2\sqrt{|K^2|}}.
\end{equation}
From the symmetry of the connection,  $\Gamma^{\sigma}_{\mu\nu}=\Gamma^{\sigma}_{\nu\mu}$, it follows $\nabla_K e_i-\nabla_{e_i} K=[K,e_i]$. Now, for any vector field belonging to the set of vector fields of class $C^{\infty}$ we have $L_X Y=[X,Y]$, this results in $\nabla_K e_i-\nabla_{e_i} K=[K,e_i]=L_K e_i=0$ by~\eqref{key} and thus in $\nabla_K e_i=\nabla_{e_i} K$, transforming~\eqref{a5} to
\begin{equation}\label{a6}
\omega_{ij}=\frac{<\nabla_{e_i} K,e_j>}{\sqrt{|K^2|}},\qquad\Omega^l=\frac{\varepsilon^{ijl}<\nabla_{e_i} K,e_j>}{2\sqrt{|K^2|}}.
\end{equation}
Since $<K,e_j>=0$ we have $<\nabla_{e_i} K,e_j>=-<K,\nabla_{e_i} e_j>$. Recalling that $\omega_{ij}$ is anti-symmetric results in 
\begin{equation}\label{a7}
\omega_{ij}=\frac{<K,\nabla_{e_j} e_i-\nabla_{e_i} e_j>}{2\sqrt{|K^2|}}=-\frac{<K,[e_i,e_j]>}{2\sqrt{|K^2|}}=-\frac{\bar{K}([e_i,e_j])}{2\sqrt{|K^2|}},
\end{equation}
where $\bar{K}$ is the one-form of $K$. Using a result from differential geometry:
\begin{equation*}
{\rm d}\omega(X,Y)=X\omega(Y)-Y\omega(X)-\omega([X,Y]),
\end{equation*}
where $X$ and $Y$ are vector fields and $\omega$ is a one-form. Let $\omega=\bar{K}$, $X=e_i$ and $Y=e_j$ we have $\bar{K}([e_i,e_j])=-{\rm d}\bar{K}(e_i,e_j)$ and the other two terms vanish: $\bar{K}(e_i)=[K,e_i]=L_K e_i=0$. Finally
\begin{equation}\label{a8}
\omega_{ij}=\frac{1}{2\sqrt{|K^2|}}~{\rm d}\bar{K}(e_i,e_j).
\end{equation}
This equation was derived in~\cite{Str} using similar analysis to the one presented here. It is straightforward to convert this equation to the one-form $\bar{\Omega}_{p}$ of $\vec{\Omega}_p=\Omega^ke_k=(\varepsilon^{ijk}\omega_{ij}/2)e_k$, as shown in~\cite{Str}
\begin{equation}\label{a9}
\bar{\Omega}_{p}=\frac{1}{2K^{2}}\ast ( \bar{K}\wedge {\rm d}\bar{K}),
\end{equation}
which is equation~\eqref{ofp}.\\

\end{document}